\documentclass[]{aa}

\usepackage{graphicx,url,twoopt,natbib,siunitx,cases,empheq}
\usepackage[version=4]{mhchem}
\usepackage{hyperref}
\hypersetup{colorlinks=true,citecolor=blue} 
\bibpunct{(}{)}{;}{a}{}{,}   

\newcommand{\avg}[1]{\left\langle #1 \right\rangle}
\newcommand{\unsim}{\mathord{\sim}}

\defcitealias{endalEvolutionRotatingStars1976}{ES}
\defcitealias{coxPrinciplesStellarStructure1968}{CG}

\begin{document}

\title{Modeling overcontact binaries}
\subtitle{I. The effect of tidal deformation}
\titlerunning{Modeling overcontact binaries I. The effect of tidal deformation}
\author{M. Fabry \and P. Marchant \and H. Sana}
\institute{Institute of Astronomy (IvS), KU Leuven, Celestijnenlaan 200D, 3001 Leuven, Belgium\\ \email{matthias.fabry@kuleuven.be} \label{ivs}}

\date{received date; accepted date}

\abstract
{
In the realm of massive stars, strong binary interaction is commonplace.
One extreme case are overcontact systems, which is expected to be part of the evolution of all stars evolving towards a merger, and is hypothesized to play a role in the formation of binary black holes.
However, important simplifications are made to model the evolution of overcontact binaries.
The deformation from tidal forces is almost always put aside, and even rotation is frequently ignored in such models.
Yet, both observation and theory show that overcontact stars are heavily tidally deformed, leaving a potentially important effect on the outer layers unaccounted for in models.
Furthermore, in eclipsing binaries where radii can be determined to high precision, it is still uncertain how large the effect of tidal deformation is on the inferred properties of stellar models.
}
{
We aim to consistently model overcontact binary stars in a one dimensional (1D) stellar evolution code.
To that end, we develop the required methodology to represent tidally distorted stars in 1D evolution codes.
}
{
Using numerical methods, we compute the structure correction factors to the 1D spherical stellar structure equations of hydrostatic equilibrium and radiative energy transfer due to the binary Roche potential, and compare them to existing results and the structure corrections of single, rotating stars.
We implement the new structure correction factors into the stellar evolution code MESA and explore several case studies.
We compare the differences between our simulations when no rotation is included, when we treat rotation using single star corrections (i.e. only accounting for centrifugal deformation) or when we use tidal deformation.
}
{
We find that ignoring rotation in deformed detached eclipsing binaries can produce a radius discrepancy of up to 5\%.
The difference between tidal and single star centrifugal distortion models is more benign at 1\%, showing that single rotating star models are a suitable approximation of tidally deformed stars in a binary system.
In overcontact configurations, we find a similar 5\% variation in surface properties as a result of tidal distortion with respect to non-rotating models, showing that it is inappropriate to model binary stars that fill their Roche lobe significantly, as non-rotating.
}
{}


\maketitle

\section{Introduction}
Massive stars ($M_{\rm initial} \gtrsim 8M_\odot$) give rise to a multitude of astrophysical phenomena on the stellar to the galactic scale.
Binarity is responsible for several of these and it is expected that the vast majority of massive stars are in binary systems that interact during their lifetime \citep{vanbeverenWROtypeStar1998,sanaBinaryInteractionDominates2012}, these effects cannot be ignored.
Furthermore, population synthesis calculations predict that about a quarter of all O stars will merge with its companion \citep{sanaBinaryInteractionDominates2012}.
Stellar mergers are thought to be responsible for strong magnetism in massive stars \citep{ferrarioOriginMagnetismUpper2009,wickramasingheMostMagneticStars2014, schneiderRejuvenationStellarMergers2016, schneiderStellarMergersOrigin2019} and are the potential source of the high rotation needed in models of long-duration gamma-ray bursts \citep{yoonEvolutionRapidlyRotating2005, yoonSingleStarProgenitors2006, aguilera-denaRelatedProgenitorModels2018} and superluminous supernovae \citep{justhamLuminousBlueVariables2014, aguilera-denaPrecollapsePropertiesSuperluminous2020}.
As mergers are preceded by an overcontact phase, understanding the properties of an overcontact model can give crucial insights in the merger process itself, and the transients associated with them, or, in a broader sense, if the merger happens at all.
\par

In the past, a significant effort has been put to observe and model the low-mass overcontact binaries, called W Ursae Majoris (W UMa) binaries.
These systems have short periods of less than a day and total masses of the order of 1$M_\odot$.
From the observational side, surveys from \citet{eggenContactBinariesII1967} and \citet{binnendijkOrbitalElementsUrsae1970} characterized the masses and mass ratio of the first identified W UMa systems, and most notably, found that the majority of them have unequal mass components.
Later, over a thousand low-mass, near-contact or overcontact binaries were found in the OGLE-I survey from light curve analysis \citep{szymanskiContactBinariesOGLEI2001}.
Extensive studies aim to characterize the population of W UMa binaries and identify systems that will merge \citep{gazeasCoBiToMProjectContact2021}. 
Recently, more and more massive overcontact binaries are identified, the most massive of which, VFTS 352 \citep{almeidaDiscoveryMassiveOvercontact2015}, has components over 30$M_\odot$.
\par

The fact that a large number of unequal overcontact binaries were observed challenged the argument from \citet{kuiperInterpretationLyraeOther1941}, who stated that overcontact stars cannot be stable, however it was shown by \citet{lucyStructureContactBinaries1968} that for the convective envelopes of W UMa binaries, this argument does not apply and they constructed an approximate model of a low mass overcontact binary.
\citet{shuStructureContactBinaries1976} expanded on these models and argued there must be significant energy transport in the innermost overcontact layers between both components.
However, the theory of overcontact models of massive stars with radiative envelopes is not yet well developed.
\par

In context of gravitational wave astrophysics, massive overcontact binaries have been proposed as progenitors of merging binary black holes (BBH) \citep{marchantNewRouteMerging2016, mandelMergingBinaryBlack2016}, as tidal synchronization leads to rapid rotation and enhanced rotational mixing \citep{deminkRotationalMixingMassive2009}.
Recently, new analysis techniques of observations allow for improved characterization of the components of a massive overcontact binary \citep{palateSpectralModellingMassive2013, abdul-masihSpectroscopicPatchModel2020, abdul-masihConstrainingOvercontactPhase2021}.
Such observations give critical constraints to evolutionary models that try to explain the evolution of overcontact systems on their way to BBH mergers.
However, to model these strongly interactive systems, important approximations are made \citep{marchantNewRouteMerging2016, menonDetailedEvolutionaryModels2021}.
Even though orbital evolution, tidal torques and mass transfer by Roche lobe (RL) overflow is usually taken care of, tidal deformation of the stellar structure from a binary companion or energy transfer in overcontact layers are ignored.
\par

Since centrifugal forces introduce a latitudinal dependence, and tidal forces an additional azimuthal one, to properly compute stellar models for single rotating or binary stars, the stellar structure equations would need to be recast and solved in 2D or 3D, respectively.
From a computational standpoint, doing full 3D stellar evolution is not feasible with current resources, so simplifying assumptions need to be made in order to remain in 1D.
\citet{kippenhahnSimpleMethodSolution1970} introduced a methodology to correct the spherical stellar structure equations to account for deformation from the spherical geometry by rotation, which was later extended by \citet{endalEvolutionRotatingStars1976}. 
This method is built on the premise that hydrostatic layers of stars fall on equipotential surfaces of the system.
For single star rotation, this method is implemented into the stellar evolution code Methods for Experiments in Stellar Astrophysics \citep[MESA,][]{paxtonModulesExperimentsStellar2011, paxtonModulesExperimentsStellar2013, paxtonModulesExperimentsStellar2015, paxtonModulesExperimentsStellar2018, paxtonModulesExperimentsStellar2019}.
Currently, these single star corrections are also called upon when evolving binary stars.
This means that while the centrifugal deformation from stellar rotation is accounted for, tidal deformation as a result of a nearby companion is not.
\par

In this first paper of a three part series, we will apply the method of \citet{kippenhahnSimpleMethodSolution1970} and \citet{endalEvolutionRotatingStars1976} to synchronized binary stars, as a first step to consistently model tidally deformed stars.
The problem of energy transfer in overcontact layers will be studied in the next paper of this series.
Still, the methods developed here are applicable to model twin overcontact binaries (binaries of mass ratio unity), where no energy is transferred between components, and also for very close detached and semi-detached systems that are tidally synchronized.
In Sect. \ref{sec:methods} we calculate the modified 1D stellar structure equations in a conservative potential and compute the structure correction factors $f_P$ and $f_T$ for a Roche binary.
In Sect. \ref{sec:og} and \ref{sec:bound}, we consider the Eddington limit and boundary conditions as appropriate to the modified equations, respectively.
Section \ref{sec:models} shows stellar models computed using the MESA code comparing currently implemented physics with the new methods in representative case studies.
Lastly, Sect. \ref{sec:conc} concludes with final remarks. 
All input files needed to reproduce the results and associated data products in this paper can be downloaded from \url{https://doi.org/10.5281/zenodo.5840879}.

\section{Methods}\label{sec:methods}
\subsection{Modifications to the spherical stellar structure equations}\label{ssec:equations}
In non-rotating stars, the equations of stellar structure and evolution are solved in a 1D space of the radius $r$ or mass coordinate $M$, and assume spherically symmetric stellar shells.
As such, the equation of mass continuity takes the form:
\begin{equation}
	\frac{\partial M}{\partial r} = 4\pi r^2 \rho,
\end{equation}
with $\rho$ the density at radius $r$ and $M=M(r)$ the total mass interior to $r$.
The equation of hydrostatic equilibrium is written as follows:
\begin{equation}
	\frac{\partial P}{\partial M} = -\frac{GM}{4\pi r^4},
\end{equation}
where $P$ is the total pressure and $G$ the gravitational constant.
Finally, in radiative regions of star, energy transport is described by:
\begin{equation}
	\frac{\partial \ln T}{\partial \ln P} = \frac{3\kappa}{16\pi ac}\frac{P}{T^4}L \equiv \nabla_{\rm rad},
\end{equation}
with $L$ the luminosity, $T$ the gas temperature, $\kappa$ the local opacity, $a$ the radiation constant and $c$ the speed of light.
However, several physical effects can break the spherical symmetry that is assumed when calculating 1D stellar evolution models.
Centrifugal forces introduce a latitudinal dependence and a binary companion introduces an additional azimuthal dependence.
Yet, if the total potential $\Psi$, which is the sum of all gravitational and rotational contributions, is conservative, \citet{kippenhahnSimpleMethodSolution1970} showed that such effects can be accounted for in 1D equations.
The only assumption that remains is the one of shellularity: We assume that stellar isobars exactly coincide with equipotential surfaces of the given geometry, therefore the total pressure $P$ is constant at constant $\Psi$.
\Citet{vonzeipelRadiativeEquilibriumRotating1924} showed that in hydrostatic equilibrium, a solidly rotating stellar shell has all its thermodynamical quantities, including $T$ and $\rho$ constant along isobars.
However, he also showed that in this case, the stellar shell cannot be in radiative thermal equilibrium.
Therefore one expects a deviation from shellularity coupled with large scale meridionial flows \citep{eddingtonCirculatingCurrentsRotating1925, sweetImportanceRotationStellar1950}.
Still, large horizontal flows are also present, and keep the rotation of isobars near to solid \citep{zahnCirculationTurbulenceRotating1992}, so that the star is kept approximately shellular.
\par
Instead of spherically symmetric stellar shells, we will write the equations of stellar structure in terms of these equipotentials surfaces, which can be characterized by a single independent variable and thus be appropriate for 1D stellar evolution.
Here we summarize the model of \citet{endalEvolutionRotatingStars1976} (henceforth \citetalias{endalEvolutionRotatingStars1976}) that cast the equations of stellar structure in 1D given a conservative potential $\Psi$.
This potential can, for example, be the Roche potential of a single rotating star (Sect. \ref{ssec:singlepot}), or of a binary star (Sect. \ref{ssec:binarypot}).
\par
We start by defining $r_\Psi$ as the equivalent radius of a sphere enclosing the same volume $V_\Psi$ enclosed by an equipotential surface:
\begin{equation}\label{eq:voleq}
	V_\Psi \equiv \frac{4\pi r_\Psi^3}{3}.
\end{equation}
Hence, as the spherical relation from radius to volume is retained and the mass density $\rho$ is assumed constant over an equipotential, the equation of continuity is unaffected:
\begin{equation}\label{eq:continuity}
	\frac{\partial M_\Psi}{\partial r_\Psi} = \rho \frac{\partial V_\Psi}{\partial r_\Psi} = 4\pi r_\Psi^2 \rho,
\end{equation}
where $M_\Psi$ is the total mass enclosed by the equipotential.
The variable $r_\Psi$ (or $M_\Psi$ in a Lagrangian prescription) will act as the new independent variable to describe stellar models.
\par
Next, we denote with $g$ the magnitude of the effective gravity at each point in space:
\begin{equation}\label{eq:grav}
	g = |\nabla\Psi| = \frac{d\Psi}{dn},
\end{equation}
$dn$ being the distance between potential surfaces $\Psi$ and $\Psi + d\Psi$.
Considering then a change in equipotential volume, we have:
\begin{align}
	dV_\Psi &= \int_{\Psi} dn dS \label{eq:dvint}\\
	&= d\Psi \int_{\Psi} g^{-1} dS = \avg{g^{-1}}S_{\Psi}d\Psi,\label{eq:dvtodpsi}
\end{align}
where $S_\Psi$ is the surface area of the equipotential, and we defined for convenience, the average of a quantity $f$ over an equipotential as
\begin{equation}\label{eq:avg}
	\avg{f} \equiv \frac{1}{S_{\Psi}}\int_{\Psi} fdS.
\end{equation}

\par

Continuing from Eq. \eqref{eq:dvtodpsi}, we have
\begin{equation}\label{eq:dpsitodm}
	d\Psi = \frac{1}{\avg{g^{-1}} S_\Psi}dV_\Psi = \frac{1}{\avg{g^{-1}}S_\Psi\rho}dM_\Psi,
\end{equation}
which, along with the statement of hydrostatic equilibrium $\nabla P = -\rho \nabla \Psi$ gives the modified momentum equation
\begin{equation}\label{eq:momentumeq}
	\frac{\partial P}{\partial M_\Psi} = -\frac{GM_\Psi}{4\pi r_{\Psi}^4}f_P,
\end{equation}
with the correction to the standard spherical equation contained in the factor
\begin{equation}\label{eq:fp}
	f_P = \frac{4\pi r_\Psi^4}{GM_\Psi S_\Psi}\frac{1}{\avg{g^{-1}}}.
\end{equation}
This factor thus measures the strength of the pressure gradient acting on a deformed shell (as a result of the non-spherical potential) with respect to a spherical shell enclosing the same volume and mass.
\par
Similarly to the mass continuity equation, the equation of energy conservation retains its spherical form as mass, radius and volume relate to each other spherically, hence
\begin{equation}
	\frac{\partial L_\Psi}{\partial M_\Psi} = \epsilon - \frac{\partial U}{\partial t} - P\frac{\partial (1/\rho)}{\partial t},
\end{equation}
where $L_\Psi$ is the energy flowing through the equipotential per unit time, $\epsilon$ is the (specific) nuclear energy generation rate, $U$ the specific internal energy, and $t$ is time.

\par
Considering lastly the equation of energy transport through radiative layers of the star, we have for the radiative flux
\begin{equation}\label{eq:vonzeipel}
	\vec{\mathcal{F}} = -\frac{4acT^3}{3\kappa\rho}\nabla T = -\frac{4acT^3}{3\kappa\rho}\frac{\partial T}{\partial \Psi}\nabla\Psi,
\end{equation}
by the chain rule.
Equation \eqref{eq:vonzeipel} is to the von Zeipel theorem \citep{vonzeipelRadiativeEquilibriumRotating1924}, stating that the radiative flux $\mathcal{F}$ is proportional to the local effective gravity $g = |\nabla \Psi|$.
Taking then the integral over an equipotential surface:
\begin{equation}\label{eq:luminosity}
	L_\Psi = \int_\Psi \vec{\mathcal{F}}\cdot\vec{dS}  = -\frac{4acT^3}{3\kappa} \avg{g^{-1}}\avg{g}S_\Psi^2\frac{\partial T}{\partial M_\Psi},
\end{equation}
using the previous results \eqref{eq:grav}, \eqref{eq:avg} and \eqref{eq:dpsitodm}.
Rewriting using \eqref{eq:momentumeq} and \eqref{eq:fp} transforms this into the usual spherical-like form of radiative energy transport:
\begin{equation}
	\frac{\partial \ln T}{\partial \ln P} = \frac{3\kappa}{16\pi acGM_\Psi}\frac{P}{T^4}L_\Psi \frac{f_T}{f_P}\label{eq:radgradient} \equiv \nabla_{\rm rad} \frac{f_T}{f_P}
\end{equation}
with the second correction factor to the spherical equations being
\begin{equation}\label{eq:ft}
	f_T = \left(\frac{4\pi r_\Psi^2}{S_\Psi}\right)^2\frac{1}{\avg{g}\avg{g^{-1}}}.
\end{equation}
Similar to $f_P$ (Eq. \eqref{eq:fp}), the correction $f_T/f_P$ measures the strength of the radiative gradient of a deformed stellar shell relative to that of a spherical shell of the same volume and enclosed mass.
\par

\subsection{Single star rotation}\label{ssec:singlepot}
A stellar shell rotating with angular velocity $\Omega$ exterior to a mass $M$ has a total potential defined by
\begin{equation}\label{eq:potrotstar}
	\Psi(r, \theta) = \Phi(r,\theta) - \frac{r^2\Omega^2\sin^2\theta}{2},
\end{equation}
where $\theta$ is the polar angle and $\Phi$ is the gravitational potential due to the mass interior to the shell.
Since stars, especially massive ones with radiative envelopes, are heavily condensed toward their centers, significant deformation from spherical symmetry is only expected in the outermost layers, justifying that the gravitational potential $\Phi$ of the star is taken as $\Phi = -\frac{GM}{r}$ of a point mass, with $M$ the total mass of the star.
\par
\citet{paxtonModulesExperimentsStellar2019} calculated analytical approximations to the corrections $f_P$ and $f_T$, and provided 
polynomial fits for these as function of the variable $\omega = \Omega/\Omega_{\rm crit} = \Omega/\sqrt{GM_\Psi/r_{\rm e}^3}$ with $r_{\rm e}$ the equatorial radius of the shell.
In appendix \ref{app:rotatingfits}, we provide new, more accurate fits without increasing the degree of the polynomial expansions. 

\subsection{Synchronized Roche binaries}\label{ssec:binarypot}
We expect that tidal torques in binary systems are very sensitive to the separation of components \citep{zahnDynamicalTideClose1975}, causing stars that significantly fill their RL to be synchronized with their orbits.
Therefore, the total potential of these systems is the well-known Roche potential and may be written in the corotating frame as follows:
\begin{equation}\label{eq:rochepot}
	\begin{split}
		\Psi(x, y, z) = \Phi(x, y, z) - \frac{G (M_1 + M_2)}{2a^3} \times \\ \left[\left(x-\frac{M_2}{M_1+M_2}a\right)^2+y^2\right],
	\end{split}
\end{equation}
where $\Phi$ is the gravitational contribution due to the two component stars, $a$ their separation and the second term represents the centrifugal contribution of the synchronous rotation around the center of mass.
We employ the same reasoning as in Sect. \ref{ssec:singlepot} to approximate the gravitational potential $\Phi$ by point mass potentials of mass $M_1$ and $M_2$, that is, the total masses of the component stars.
Finally, defining the mass-ratio $q = M_2/M_1$, normalizing all distances $r \rightarrow r/a$ and shifting $\Psi \rightarrow \frac{\Psi a}{GM_1} - \frac{q^2M_1}{1+q}$, we arrive at the dimensionless form of the potential in spherical coordinates:
\begin{equation}\label{eq:rochepotdimless}
	\Psi(r, \theta, \phi) = -\frac{1}{r}-q\left(\frac{1}{r'}-r\sin\theta\cos\phi\right)-\frac{1+q}{2}r^2\sin^2\theta,
\end{equation}
where $r^2 = x^2 + y^2 + z^2,\, r\sin\theta\cos\phi = x,\, r\sin\theta\sin\phi = y$ and $r\cos\theta=z$ relate to the Cartesian frame, and $r' = \sqrt{1-2r\sin\theta\cos\phi+r^2}$.
In Fig. \ref{fig:roche}, we show an illustration of the Roche geometry in the equatorial plane and highlight the colinear Lagrangian point equipotentials.
\par

\subsection{Overcontact shells}\label{ssec:split}
\begin{figure}
	\centering
	\includegraphics[width=\columnwidth]{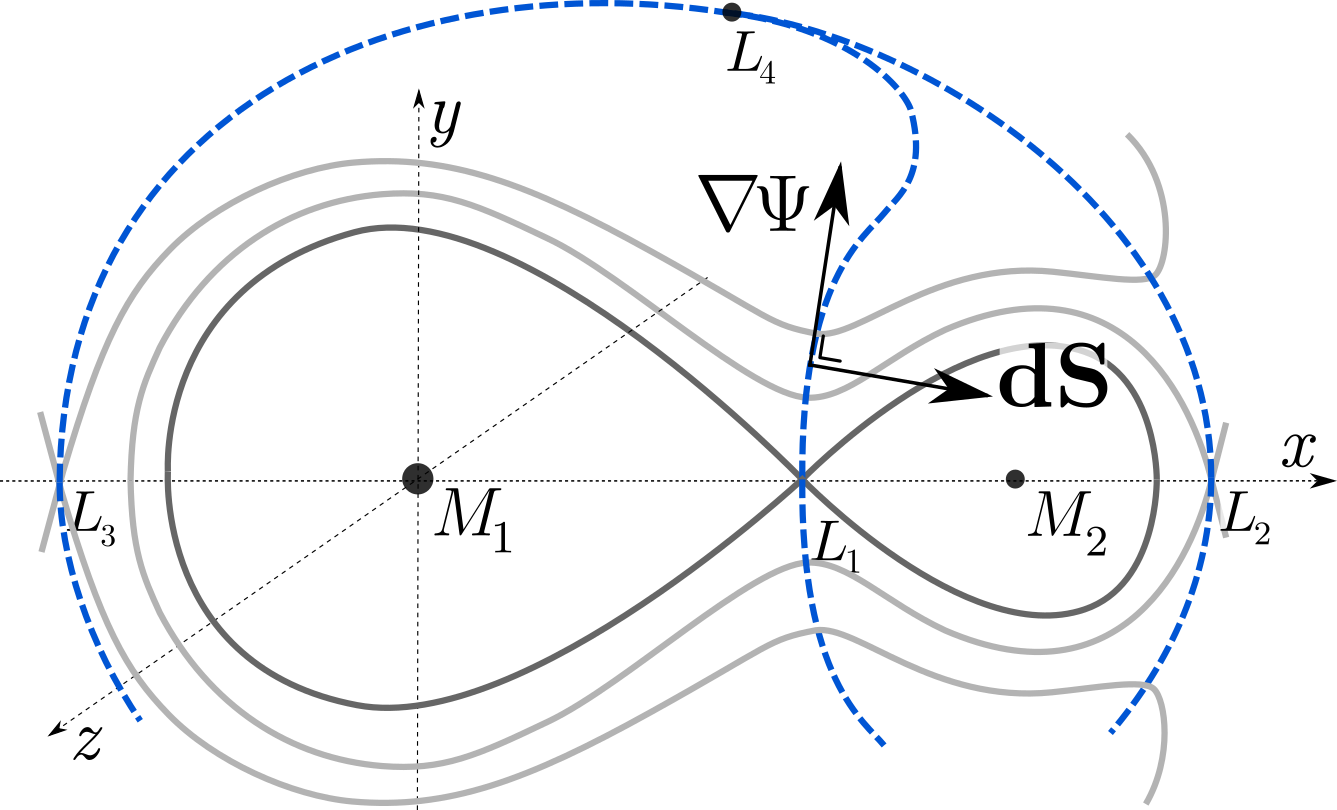}
	\caption{Illustration of the Roche geometry in the equatorial plane. It shows the $L_1, L_2$ and $L_3$ equipotentials (gray), as well as the splitting surfaces (blue dashed). The splitting surfaces have their normals always perpendicular to the potential gradient.}
	\label{fig:roche}
\end{figure}

In the context of 1D evolution models, we have to distinguish the two stellar components within the geometry of a Roche binary.
For equipotential shells lying within their respective RLs, this is trivial, as such layers are physically separated, but shells in overcontact are shared between the two stars.
We therefore construct three ``splitting surfaces'', one through each of the colinear Lagrangian points $L_1, L_2$ and $L_3$, separating the Roche geometry in three main parts.
This will ensure distinction between the two components, as well as either from regions beyond the outer Lagrangian points $L_2$ and $L_3$, from which outflows are expected if any component overflows them.
We construct them in such a way that the condition
\begin{equation}\label{eq:surfconstruction}
	\nabla \Psi \cdot \vec{dS} = 0
\end{equation}
holds on all points of these splitting surfaces (see Fig. \ref{fig:roche}, in blue dashed lines), such that the surface normal is always perpendicular to the gradient of the potential \eqref{eq:rochepotdimless}.
Note that by our construction of the splitting surfaces, the scalar product $\vec{dn}\cdot\vec{dS}$ is zero.
This ensures that Eq. \eqref{eq:dvint} still holds, and if the overcontact layers are shellular, there is no net flux of radiative energy between the components following Eq. \eqref{eq:vonzeipel}
However, since the condition of shellularity cannot be held exactly, there will exist a flow $L_{\Psi, \rm trans} = \int_{d\Psi} \vec{\mathcal{F}}\cdot\vec{dS_{\rm split}}$ across equipotentials, representing energy transfer from one component to the other.
In this work, we assume this contribution is zero and defer the modeling of energy transfer to later work.
We note also that in the case $q \ne 1$, the $L_1$ splitting surface is not a vertical plane through $L_1$, which differentiates our results of equipotential volumes and surface areas from those of \citet{mochnackiAccurateIntegrationsRoche1984} and \citet{marchantRoleMassTransfer2021}.
\par

\subsection{Numerical calculations}\label{ssec:num}
In order to use the modified stellar structure equations in a stellar evolution code, we require for each shell in the geometry the quantities $V_\Psi$, $S_\Psi$, $\avg{g}$ and $\avg{g^{-1}}$ which specify $r_\Psi, f_P$ and $f_T$ of Eqs. \eqref{eq:voleq}, \eqref{eq:fp} and \eqref{eq:ft}.     
Additionally, the specific moment of inertia $i_{\rm rot}$ is needed to calculate the rotational velocity of the equipotential shell from the angular momentum: $\Omega_{\Psi} = j_{\rm rot}/i_{\rm rot}$.
We approach this problem fully numerically.
First of all we note that the problem is symmetric under $y \rightarrow -y$ and $z \rightarrow -z$, so we need only explicitly do calculations in the quadrant of positive $y$ and $z$.
\par
Given a set of mass ratios $q$ and target equipotential values $\Psi_0$, we calculate, in a spherical coordinate system on rays of constant $(\theta, \phi)$, the points $r = r(\theta, \phi)$ that lie on the corresponding equipotential surface, such that:
\begin{equation}\label{eq:equiroot}
	\Psi_0 - \Psi(r, \theta, \phi) = 0.
\end{equation}
When solving Eq. \eqref{eq:equiroot}, care has to be taken in order to avoid roots beyond the splitting surfaces (Sect. \ref{ssec:split}), as these belong to a part the geometry not associated to the considered component star, and would instead belong to the other component, or to the region beyond the outer Lagrangian points.
We therefore restrict the solver to values $r$ smaller than the value $r_{\rm split}(\theta, \phi)$ on the splitting surface that the ray $(\theta, \phi)$ crosses.
If it finds no roots in the interval $(0, r_{\rm split})$, $r_{\rm split}$ is recorded instead as the splitting surfaces mark the maximal extent of the components.
\par
Next, for these points $r(\theta, \phi)$, we numerically integrate the quantities:
\begin{align}
	V_\Psi &= \int_\Psi \frac{r^3}{3}\sin\theta d\theta d\phi,\label{eq:vpsi}\\
	S_\Psi &= \int_\Psi \frac{1}{\hat{n}\cdot\hat{r}}r^2\sin\theta d\theta d\phi, \\
	\avg{g} &= \frac{1}{S_\Psi} \int_\Psi \frac{g}{\hat{n}\cdot\hat{r}}r^2\sin\theta d\theta d\phi, \\
	\avg{g^{-1}} &= \frac{1}{S_\Psi} \int_\Psi \frac{1}{g \hat{n}\cdot\hat{r}}r^2\sin\theta d\theta d\phi, \\
	i_{\rm rot} &= \frac{\int_\Psi (x^2+y^2) dm_\Psi}{\int_\Psi dm_\Psi} = \frac{\int_\Psi r^2\sin^2\theta g^{-1}d\Psi dS}{\avg{g^{-1}} S_\Psi d\Psi} \notag \\
	&=\frac{1}{S_\Psi \avg{g^{-1}}} \int_\Psi \frac{ r^2\sin^2\theta}{g\hat{n}\cdot\hat{r}}r^2\sin\theta d\theta d\phi.\label{eq:irot},
\end{align}
where the $\hat{n}\cdot\hat{r} = \left.\frac{d\Psi}{dr}\middle/|\nabla\Psi|\right.$ and we used the notation $dm_\Psi = \rho dn dS$ to signify a small mass element within the equipotential shell, such that $dM_\Psi = \int_\Psi dm_\Psi$.
From these, $f_P$ and $f_T$ can be computed for each equipotential shell with Eqs. \eqref{eq:fp} and \eqref{eq:ft}.

\par

\subsection{Integration results}\label{ssec:intresults}
We consider the component of a Roche binary at the origin with mass $M_1$ and a companion with mass $M_2$, and computed all necessary integrals of Eqs. \eqref{eq:vpsi}-\eqref{eq:irot} in the two-dimensional parameter space of mass ratio $q \equiv M_2/M_1$ and Roche equipotentials $\Psi_0$.
We emphasize here that we calculate solely the properties of the component with mass $M_1$ and that the properties of the companion $M_2$ can be obtained by inverting the mass ratio.
We sampled 280 equally spaced mass ratios in logarithmic space in the interval $\log q \in [-7, 7]$, as well as the equal mass ratio $\log q = 0$ case while in potential space, we sampled 158 values in several logarithmic ranges from $\Psi_0 = 50\Psi_{L_1}$ to $\Psi_0 = \frac{1}{2}(\Psi_{L_{\rm out}} + \Psi_{L_4})$ of different densities.
With $L_{\rm out}$ we denote the outer Lagrangian point of the considered component of the binary (always having $x_{L_{\rm out}} < 0$, which is $L_3$ if $q \leq 1$ and $L_2$ if $q \geq 1$).
As quantities exhibit important variations around the overflow potentials $\Psi_{L_1}$ and $\Psi_{L_{\rm out}}$, those will be the regions we sample more densely.
Note that $\Psi_{L_1} < \Psi_{L_2} \leq \Psi_{L_3} < \Psi_{L_4} = \Psi_{L_5}$ as in our convention $\Psi$ is strictly negative.
\par
\begin{figure}
	\centering
	\includegraphics[width=\columnwidth]{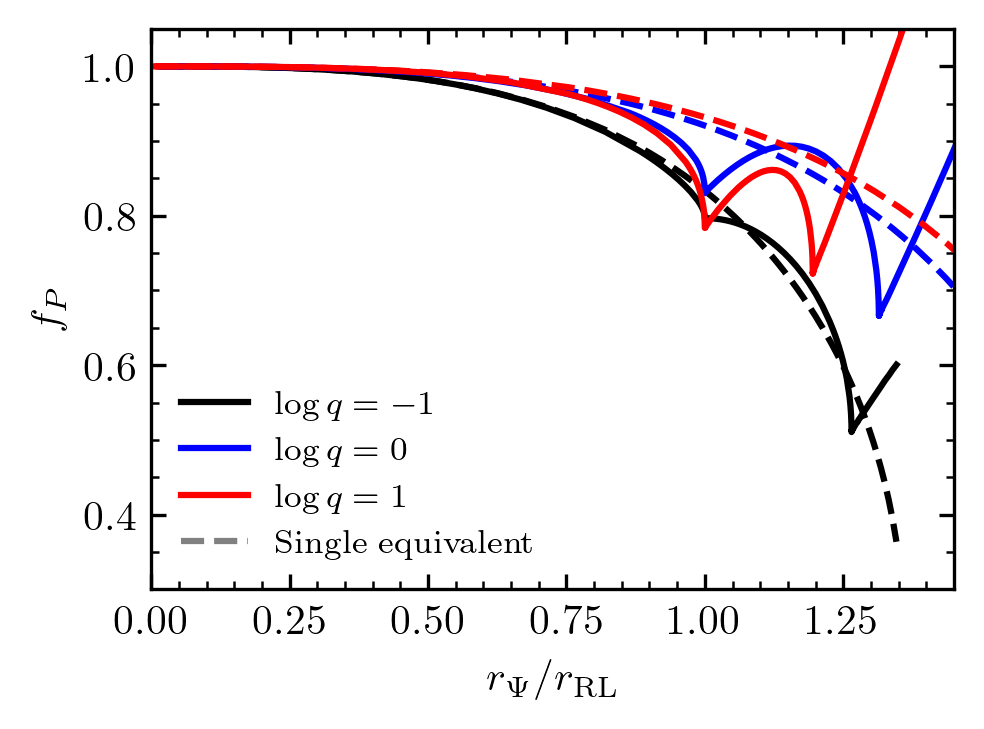}
	\caption{Structure correction factor $f_P$ as function of fractional RL radius $r_\Psi/r_{\rm RL}$ for various mass ratios. The discontinuities in all graphs correspond to the crossing of the Lagrangian points $L_1$ (at $r_\Psi/r_{\rm RL}=1$, by construction) and $L_{\rm out}$ at larger radii. For reference, the equivalent single rotating star inferred $f_P$ is overplotted in dashed lines.}
	\label{fig:fp_r_super}
\end{figure}

Figure \ref{fig:fp_r_super} shows the integration results of $f_P$ for various mass ratios as function of their radius relative to the RL radius $r_\Psi/r_{\rm RL}$, where we defined $r_{\Psi_{L_1}} \equiv r_{\rm RL}$.
We observe discontinuities at the Lagrangian overflow points, which is a result of the splitting surfaces limiting the considered volume to one component (Fig. \ref{fig:roche}).
For comparison with the single rotating star deformation model, we overplot the equivalent $f_P$ values as if the binary star was interpreted as a single rotating star with the same volume and rotational velocity (see Appendix \ref{app:rotating-roche} for details).
\par

\subsection{Comparison to literature}
\begin{figure}
	\centering
	\includegraphics[width=\columnwidth]{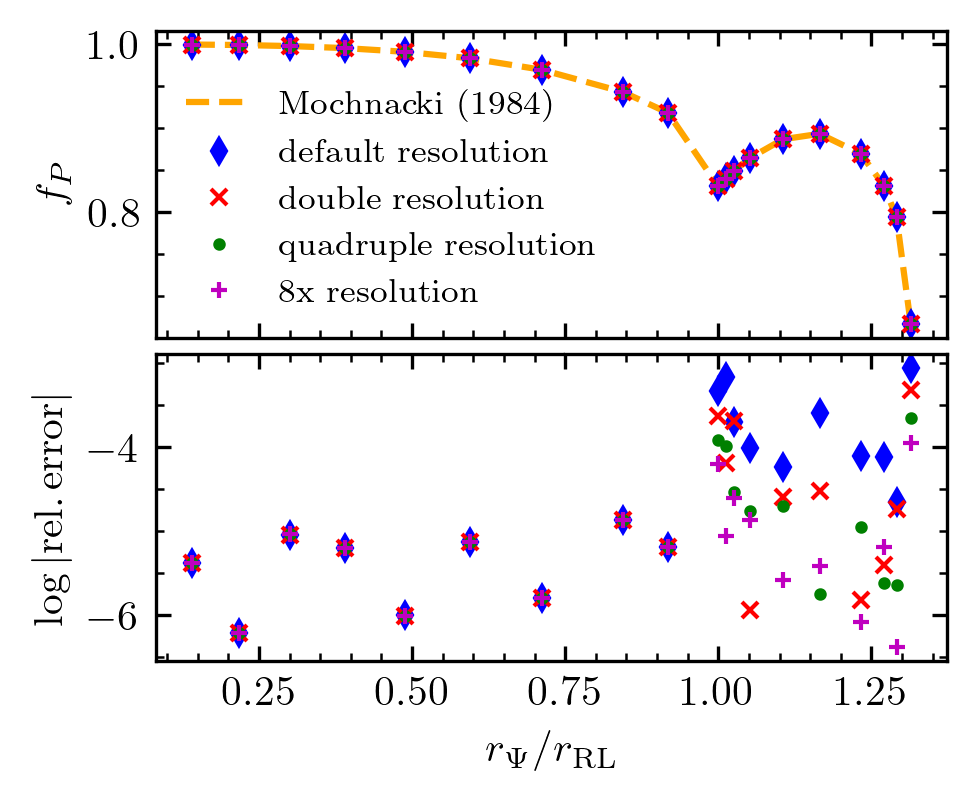}
	\caption{Comparison of the calculations of the structure correction factor $f_P$ from \citet{mochnackiAccurateIntegrationsRoche1984} and those in this work, as function of the fractional RL radius $r_\Psi/r_{\rm RL}$. 
		We performed a resolution convergence test with double, quadruple and octuple the default spacial resolution, and conclude our integrations have well converged at an eight-fold resolution increase.}
	\label{fig:mochcomp}
\end{figure}

As verification of our integrations, we compare our results to calculations performed by \citet{mochnackiAccurateIntegrationsRoche1984}.
As this study considered vertical splitting surfaces through $L_1$ at all mass ratio's, for overcontact layers we are limited to comparing the mass ratio unity case.
Figure \ref{fig:mochcomp} shows excellent agreement between our $f_P$ values and those computed via Eq. \eqref{eq:fp} using the results from \citet{mochnackiAccurateIntegrationsRoche1984}, with the relative difference being less than about 1 part in $10^3$.
A resolution convergence test of our integrations show our calculations converge to the $10^{-5}$ level at 8 times our original default resolution.
The final computations of all integrals for use in stellar evolution instruments are done with this eight-fold resolution increase, and corresponds to dividing the interval $\phi=[0, \pi]$ in $n_\phi=5280$ parts and $\cos\theta = [0, 1]$ in $n_\theta=n_\phi/2$ parts so that $\Delta\phi = \pi/n_\phi$ and $\Delta\cos\theta = 2/n_\phi$.
\par

\section{The $\Omega\Gamma$ limit}\label{sec:og}
Thanks to radiation pressure of the photon flux generated in the core, a star gains radiative support against its self gravity.
In the situation where these effects cancel exactly, the star has reached the Eddington limit, which is stated as (for a single, non-rotating and thus spherical star):
\begin{equation}
	L = L_{\rm Edd} \equiv \frac{4\pi cGM}{\kappa}.
\end{equation}
Defining then the Eddington factor $\Gamma$ as:
\begin{equation}
	\Gamma \equiv \frac{L_{\rm rad}}{L_{\rm Edd}},
\end{equation}
with $L_{\rm rad}$ the radiative luminosity, equal to $L$ in a radiative envelope, $\Gamma=1$ is then an equivalent statement of the Eddington limit.
If $\Gamma > 1$, the sum of radiative and gravitational acceleration is directed outward.
In the interior of the star this can in principle be compensated by an inversion of the gas pressure \citep{jossCriticalLuminosityStellar1973}, but if a star approaches $\Gamma = 1$ at the surface, strong outflows are expected to develop \citep{grafenerEddingtonFactorKey2011}.
\par

Adding a centrifugal contribution to the force balance, we have:
\begin{equation}
	\vec{g_{\rm tot}} = \vec{g_{\rm grav}} + \vec{g_{\rm rot}} + \vec{g_{\rm rad}} = \vec{g} + \vec{g_{\rm rad}} = 0,
\end{equation}
where we write $\vec{g} = \nabla \Psi$ as the effective gravity.
\citet{langerEddingtonLimitRotating1997} then considered the Eddington limit on the equator of a rotating (but non-deformed) star, and, assuming that the radiation field is isotropic, the radiative flux is $\mathcal{F} = L/4\pi R^2$,
so that the force balance reduces to
\begin{equation}\label{eq:langer_crit}
	1-\omega^2 - \Gamma = 0,
\end{equation}
where we have defined $\omega = \Omega/\Omega_{\rm crit, class}$ as the fractional classical critical rotation rate $\Omega_{\rm crit, class} = \sqrt{\frac{GM}{R^3}}$.
It is important to make the distinction here as the formal critical velocity is lowered to:
\begin{equation}\label{eq:langerv}
	v_{\rm crit} = \sqrt{\frac{GM}{R}\left(1-\Gamma\right)} = \sqrt{v_{\rm crit, class}(1-\Gamma)}.
\end{equation}
Therefore, the closer a star is to the Eddington limit, the lower the critical velocity will be.
\par
\citet{maederStellarEvolutionRotation2000} later noted that this treatment does not take von Zeipel's theorem into account, namely that the radiative flux is dependent on local effective gravity, Eq. \eqref{eq:vonzeipel}. 
Expressing this for a rotating star to lowest order in rotational velocity $\Omega$ gives
\begin{equation}\label{eq:GObalance}
	\vec{g}_{\rm tot} = \vec{g}\left(1-\Gamma_\Omega(\vartheta)\right),
\end{equation}
where $\vartheta$ signifies a polar dependence and $\Gamma_\Omega \equiv \Gamma \times f(\Omega, \vartheta)$ is now a rotation dependent Eddington factor.
Since the strength of the radiative flux is a function of the effective gravity through the von Zeipel effect, \citet{maederStellarEvolutionRotation2000} showed that, neglecting the $\vartheta$ dependence, Eq. \eqref{eq:GObalance} bifurcates when $\Gamma = 0.639$.
Below this, $\vec{g} = 0$ is the only solution, so that, independent of luminosity, the break up velocity is the classical velocity $v_{\rm crit,class} = \sqrt{\frac{GM}{r_{\rm e}}}$, with $r_e$ the equatorial radius of the rotating star.
Above the bifurcation point, both $\vec{g} = 0$ and importantly, 
\begin{equation}\label{eq:maederbalance}
	1-\Gamma_\Omega = 0,
\end{equation}
determine critical velocities, the second of which is lower than the classical critical velocity.
The lowest of these two velocities then determines the physical break-up velocity of a rotating star.
\par
We further expand on the notion of break-up limits by considering the case of deformed stars due to a conservative potential.
As a starting point, we consider the same criterion of zero net acceleration $\vec{g_{\rm tot}} = 0$ as the stability boundary.
The contributions to this acceleration are the effective gravity $\vec{g}$ as a result of the potential, and the radiative acceleration.
Using von Zeipel's expression for radiative flux (Eq. \eqref{eq:vonzeipel}), and combining with the equation of hydrostatic equilibrium results in:
\begin{equation}\label{eq:radaccel}
	\vec{g_{\rm rad}} = -\frac{4aT^3}{3} \frac{dT}{dP} \vec{g},
\end{equation}
which, upon using Eq. \eqref{eq:radgradient} as the radiative gradient in the non-spherical geometry, gives the stability criterion:
\begin{equation}\label{eq:radequil}
	\vec{g}\left(1-\Gamma\frac{f_T}{f_P}\right) = 0.
\end{equation}
We recognize however that von Zeipel's law of radiation can break down close to critical velocity due to significant baroclinicity, that is, departure from shellularity (see for example \citealp{espinosalaraGravityDarkeningRotating2011}).
Continuing on our assumption of shellularity, Eq. \eqref{eq:radequil} is of a very similar form as that of \citet{maederStellarEvolutionRotation2000}, namely $\vec{g}(1-\Gamma_\Psi) = 0$, where $\Gamma_\Psi = \Gamma\times f$ is again a product of the classical Eddington factor with a function dependent on the geometry.
For the single rotating star deformation $f=f(\omega)$ dependent on the fractional critical rotation rate, while in the tidal deformation case $f=f(r_\psi/r_{\rm RL}, q)$ dependent on the mass ratio and degree of Roche filling.
In Eq. \eqref{eq:radequil}, the classical solution $\vec{g} = 0$ is present of course, where the effective gravity vanishes as a result of rotational support only.
As above, in this case there is no corresponding Eddington luminosity as it is independent of any radiative acceleration due to the Von Zeipel effect.
The other solution is given by:
\begin{equation}\label{eq:ftfpbalance}
	1-\Gamma\frac{f_T}{f_P} = 0,
\end{equation}
which translates to a modified expression of the Eddington luminosity:
\begin{equation}
	L_{\rm Edd,\Psi} = L_{\rm Edd}\frac{f_P}{f_T},
\end{equation}
or the Eddington factor \citep[see also][]{sanyalMassiveMainsequenceStars2015}:
\begin{equation}
	\Gamma_\Psi = \frac{L_\Psi}{L_{\rm Edd, \Psi}} = \frac{\kappa L_\Psi}{4\pi cGM_\Psi}\frac{f_T}{f_P}.
\end{equation}
These expressions, like the equations derived in Sect. \ref{ssec:equations}, are applicable in any conservative potential $\Psi$.
\par
\begin{figure}
	\centering
	\includegraphics{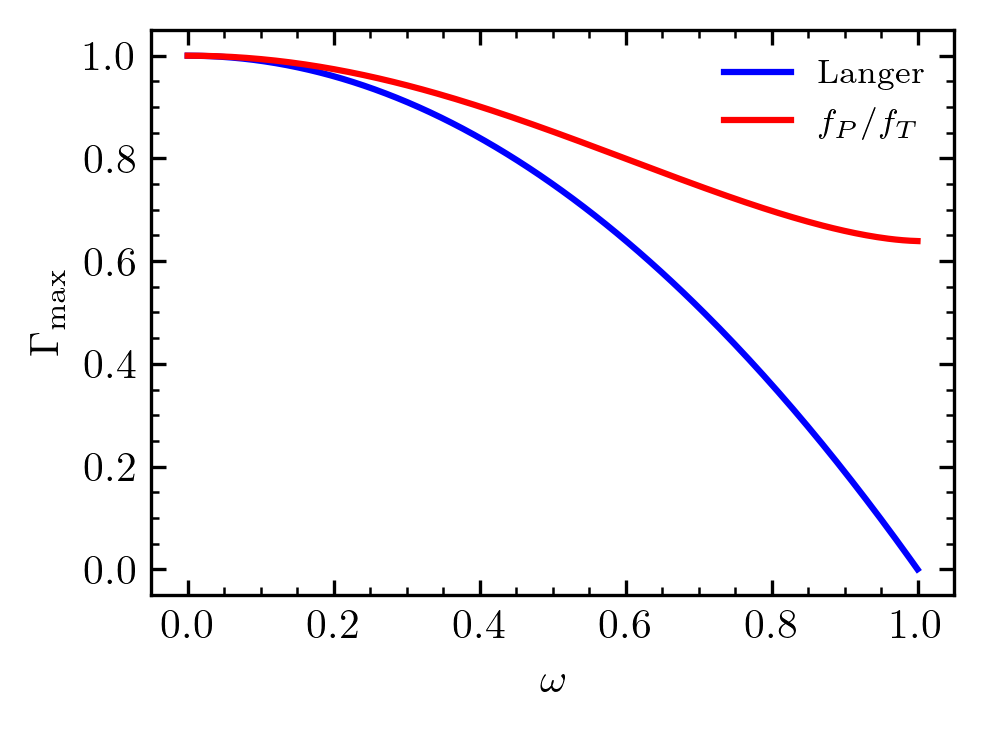}
	\caption{Maximal Eddington factor $\Gamma_{\rm max}$ a rotating star can have as function of fractional classical critical rotation rate $\omega = \Omega/\Omega_{\rm crit, class}$. Our model of shellular stars has a maximal reduction to $\Gamma_{\rm max} = 0.639$.}
	\label{fig:gamma_omega_single}
\end{figure}
\begin{figure}
	\centering
	\includegraphics{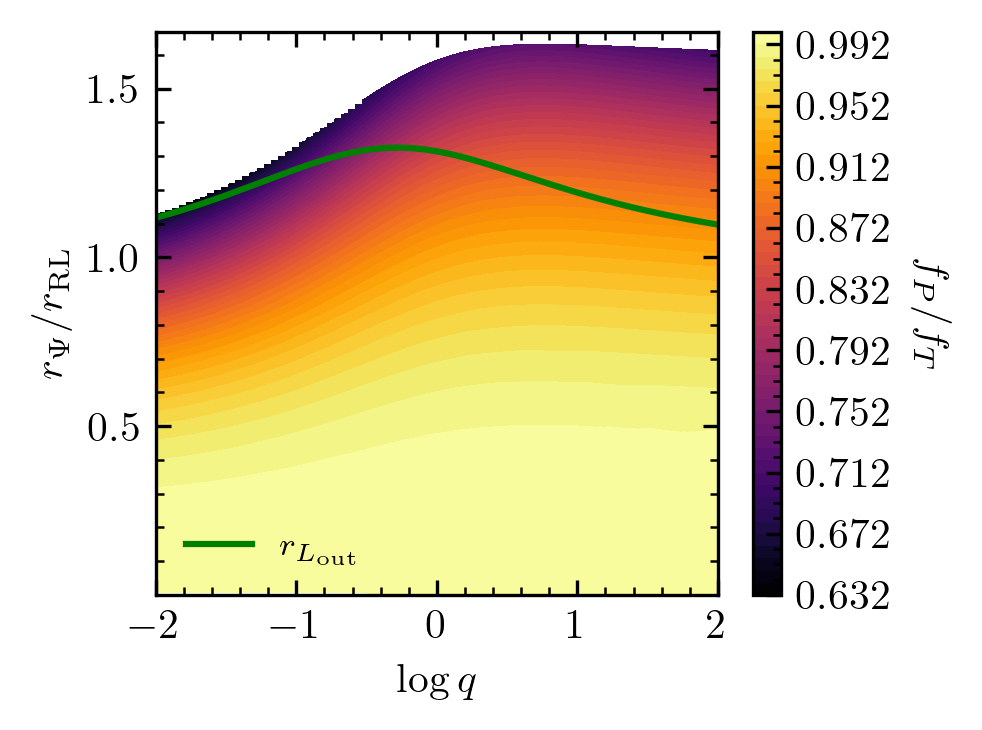}
	\caption{Reduction $f_P/f_T$ of the maximal Eddington factor as function of mass ratio $q$ and fractional overflow radius $r_\Psi/r_{\rm RL}$. The equivalent radius of a star filling up to its outer Lagrangian point in marked with the green line.}
	\label{fig:fpftratio_tidal}
\end{figure}

As criteria for instability, Eqs. \eqref{eq:langer_crit}, \eqref{eq:maederbalance} and \eqref{eq:ftfpbalance} all express a departure of the Eddington limit of a star from the classical Eddington limit $\Gamma = 1$.
In the model developed here, it amounts to a reduction of the maximal Eddington factor by the ratio $f_P/f_T$, which is smaller than one in both the single rotating star or synchronized binary case.
In Fig. \ref{fig:gamma_omega_single}, we show the comparison of this reduction from the results of \citet{langerEddingtonLimitRotating1997} to this work in the case of a single rotating star.
Moreover, consistent to \citet{maederStellarEvolutionRotation2000}, we find a bifurcation in the critical rotation velocity at at $\Gamma = 0.639$.
Below this number, only the classical critical velocity $\omega = 1$ will make the star unstable, while above it, this velocity is reduced.
The case of a synchronized Roche binary is shown in Fig. \ref{fig:fpftratio_tidal}.
For example, an equal mass ratio binary overflowing to its outer Lagrangian point has its Eddington limit reduced to about 84\% its classical value.
\par


\section{Atmospheric boundary conditions}\label{sec:bound}
The equations of stellar structure must be supplied with appropriate boundary conditions (BCs) at the center and atmosphere of the star.
In the center, the conditions are such that $r=M=L=0$, which transform to $r_\Psi = M_\Psi = L_\Psi = 0$ in our equipotential shell model.
At the surface, an atmosphere model is integrated to supply the pressure $P_{\rm surf}$ and temperature $T_{\rm surf}$.
A common choice is the so called gray Eddington atmosphere, which is a plane parallel model integrated from $\tau=0$ to some predetermined atmospheric optical depth $\tau_{\rm surf}$, and is presented in, for example, \citet{coxPrinciplesStellarStructure1968} (henceforth \citetalias{coxPrinciplesStellarStructure1968}).
\par
In spherical stars, the surface optical depth $\tau_{\rm surf}$ is reached at the same physical perpendicular depth $z$ at all points in the star.
For deformed shellular stars however, the optical depth $\tau$ across an equipotential depends on local effective gravity, as follows:
\begin{equation}\label{eq:opdepth}
	d\tau = \kappa\rho dz = -\kappa \rho \frac{dn}{d\Psi}d\Psi = -\kappa\rho g^{-1}d\Psi,
\end{equation}
with $\kappa$ and $\rho$ the opacity and density of the equipotential, respectively.
Consequently, equipotential surfaces do not coincide with surfaces of constant optical depth.
For example, in single rotating stars, a fixed $\tau_{\rm surf}$ is reached at a deeper equipotential on the poles compared to on the equator, resulting in a different observed effective temperature between those regions.
The same is true for highly deformed binary stars like overcontact systems.
The atmosphere properties output by stellar evolution codes should therefore also take into account this effect of position dependent atmospheric depth.
\par
Within the treatment of deformed stars in a conservative potential, we proceed as follows.
We start by defining the global effective temperature $T_{\rm eff, \Psi}$ of an equipotential by the law of Stefan-Boltzmann:
\begin{equation}\label{eq:globalteff}
	L_\Psi = \int_\Psi \mathcal{F}dS \equiv S_\Psi\sigma T_{\rm eff, \Psi}^4,
\end{equation}
that is, the global effective temperature of an equipotential shell is that of a black body radiating a luminosity $L_\Psi$ over an area $S_\Psi$, which need not be spherical.
Next, using the equation of radiative flux Eq. \eqref{eq:vonzeipel}, we can define a local effective temperature $T_{\rm eff, l}$ that varies across the equipotential:
\begin{equation}\label{eq:localteff}
	\mathcal{F} = \frac{g}{\avg{g}}\sigma T_{\rm eff, \Psi}^4 = \sigma T_{\rm eff, l}^4.
\end{equation}
\par
We now consider a gray, plane parallel, Eddington-approximated atmosphere to construct appropriate BC.s
At each point on the surface equipotential, the temperature profile can be written as \citepalias{coxPrinciplesStellarStructure1968}:
\begin{equation}\label{eq:tempprofile}
	T^4(\tau) = \frac{1}{2}T_{\rm eff, l}^4\left(1+\frac{3}{2}\tau\right).
\end{equation}
This expression is derived using two standard assumptions that can be similarly applied to the case of deformed stars.
First, it is assumed that the radiation pressure at all optical depths can be written as $P_r(\tau) = \frac{1}{3}aT^4(\tau)$,
and second, the intensity of radiation on top of the atmosphere is assumed to be isotropic in the outward directions, meaning we ignore limb darkening, so that the pressure at the $\tau=0$ surface is computed to be \citepalias{coxPrinciplesStellarStructure1968}:
\begin{equation}\label{eq:outerpres}
	P_r(0) = \frac{2\mathcal{F}}{3c}.
\end{equation}
\par

The first BC we consider constrains the temperature $T_{\rm surf}$ of the outermost boundary by requiring that $T_{\rm surf} = T_{\rm eff, \Psi}$.
To find the second BC on the surface pressure $P_{\rm surf}$, we consider the equation of hydrostatic equilibrium:
\begin{equation}
	\frac{dP}{d\tau} = \frac{g}{\kappa}.
\end{equation}
Taking a point on the surface equipotential where $T_{\rm eff, l} = T_{\rm eff, \Psi}$, then by Eq. \eqref{eq:localteff} we have that $g = \avg{g}$.
As we also require that $T_{\rm surf} = T_{\rm eff, \Psi}$, Eq. \eqref{eq:tempprofile} implies that the surface optical depth is $\tau_{\rm surf} = 2/3$.
The surface pressure can then be integrated at this point of the equipotential surface by assuming a thin atmosphere such that $g$ and $\kappa$ are constants.
\begin{equation}\label{eq:photpressure}
	P_p = \int_0^{\tau_{\rm surf}}\frac{\avg{g}}{\kappa}d\tau + P_r(0)= \frac{2}{3}\frac{4\pi GM_\Psi}{\kappa S_\Psi}\frac{f_P}{f_T} + \frac{2}{3}\frac{\sigma T_{\rm surf}^4}{c},
\end{equation}
where we substituted $\avg{g}$ with the definitions of $f_P$ and $f_T$ in Eqs. \eqref{eq:fp} and \eqref{eq:ft}, respectively.
\par
In summary, the atmospheric BCs are the system of equations:
\begin{empheq}[left=\empheqlbrace]{align}
	P_{\rm surf} &= \frac{2}{3}\frac{4\pi GM_\Psi}{\kappa(T_{\rm surf}, P_{\rm surf})S_\Psi}\frac{f_P}{f_T} + \frac{2}{3}\frac{\sigma T_{\rm surf}^4}{c},\label{eq:pressbound}\\
	T_{\rm surf} &= \sqrt[4]{\frac{L_\Psi}{\sigma S_\Psi}}.\label{eq:tempbound}
\end{empheq}
We emphasize that this surface temperature $T_{\rm surf}$ of the stellar models is not the local effective temperature across the whole surface equipotential, the variations of which can be observed.
However, it is possible to relate the surface temperature output by stellar models using these BCs to an observed local effective temperature at any point on the surface equipotential by applying Eq. \eqref{eq:localteff}.
Finally, we mention that from now on, whenever the effective temperature is referenced, we mean the global effective temperature as defined by Eq. \eqref{eq:globalteff}.
\par
\section{Stellar model comparison}\label{sec:models}
We assess the effect of including tidal distortion on the structure of the star as well as the modified boundary conditions by computing stellar evolution models using the 1D MESA code, version 15140, with the \texttt{mesasdk-x86\_64-macos-21.2.1} SDK.
Starting at the zero age main sequence (ZAMS), we compare the evolution of stars in three cases: (1) a $5M_{\odot}$ single rotating star, (2) a $5M_{\odot}$ detached star in orbit with a compact object, and (3) a $32M_{\odot}$ twin binary in overcontact.
\par
The next subsection lists the physical assumptions made when computing stellar models.
Following subsections detail and discuss the results of the respective case studies.

\subsection{Physical ingredients}
\subsubsection{Mixing, microphysics and stellar winds}
In all stellar models, we use the Ledoux criterion \citep{ledouxStellarModelsConvection1947} for determining convective regions.
Within such regions, convection itself is modeled via the mixing length theory of \citet{bohm-vitenseUberWasserstoffkonvektionszoneSternen1958} as described by \citetalias{coxPrinciplesStellarStructure1968}, with the mixing length parameter $\alpha = 2$.
We use efficient semiconvective mixing \citep{langerSemiconvectiveDiffusionEnergy1983} with $\alpha_{\rm sc} = 100$ as proposed by \citet{schootemeijerConstrainingMixingMassive2019}.
Thermohaline mixing is applied as prescribed in \citet{kippenhahnTimeScaleThermohaline1980}, with an efficiency of $\alpha_{\rm th} = 1$.
During the main sequence, the convective core is allowed to overshoot into the radiative envelope. 
We use a step overshoot as calibrated by \citet{brottRotatingMassiveMainsequence2011} with $f = 0.345, f_0 = 0.01$, meaning that the diffusion coefficient is taken as constant from $f_0$ pressure scale heights into the convective boundary up to $f-f_0$ pressure scale heights out of the boundary.
All other extra mixing processes, including rotational mixing from, for example, Eddington-Sweet type circulations are ignored in these sample models so as to isolate the effects of including the different forms of rotational deformation.
\par
In terms of microphysics, we use the basic nuclear network consisting of $\ce{^{1}H, ^{3}He, ^{4}He, ^{12}C, ^{14}N, ^{16}O, ^{20}Ne}$ and $\ce{^{24}Mg}$, which is sufficient to follow the main sequence evolution \citep{paxtonModulesExperimentsStellar2011} and the reaction rates are taken from the JINA library \citep{cyburtJINAREACLIBDatabase2010}.
MESA uses a blend of different equation of states (EOS) from \citet{saumonEquationStateLowMass1995}, \citet{timmesAccuracyConsistencySpeed2000}, \citet{rogersUpdatedExpandedOPAL2002} and \citet{potekhinThermodynamicFunctionsDense2010}, as described in \citet{paxtonModulesExperimentsStellar2019}.
Radiative opacities are similarly determined from a blend of opacity tables from \citet{iglesiasUpdatedOpalOpacities1996} and \citet{fergusonLowTemperatureOpacities2005}.
The solar metallicity $Z_{\odot} = 0.0142$ as well as the relative metal fractions from are taken from \citet{asplundChemicalCompositionSun2009}.
\par
Wind mass loss of individual stars is included following the prescription of \citet{brottRotatingMassiveMainsequence2011}.
As we focus on main sequence evolution, where the surface hydrogen fraction is $X>0.7$, this method takes rates from \citet{vinkMasslossPredictionsStars2001} when the effective temperature is higher than the iron bi-stability jump (as calibrated by \citealt{vinkMasslossPredictionsStars2001}).
For effective temperatures below the bi-stability, the mass loss rate is computed as the maximum of the rates of \citet{vinkMasslossPredictionsStars2001} and \citet{nieuwenhuijzenAtmosphericAccelerationsStability1995}, where the latter is scaled with the same metallicity factor $(Z/Z_\odot)^{0.85}$ like the rates of \citet{vinkMasslossPredictionsStars2001}.

\subsubsection{Rotation and binary physics}
At all times during the evolution of all models, angular momentum is continuously diffused throughout the whole star so as to enforce rigid rotation during the evolution.
This is, like omitting rotational mixing, a way to reduce differences between the different rotational distortion models.
When modeling single star rotation, the structure correction factors $f_P$ and $f_T$ that we use are those from appendix \ref{app:rotatingfits}.
In binary models, the angular velocity is always synchronized to the orbital period, so that the Roche potential \eqref{eq:rochepotdimless} applies and the timescale of this synchronization is the orbital period itself.
The structure corrections $f_P$ and $f_T$ are determined by interpolating the results of Sect. \ref{ssec:intresults} on a grid of mass ratio and fractional RL radius $(q = M_2/M_1, r_\Psi/r_{\rm RL})$, where the RL sizes are determined according to the Eggleton approximation \citep{eggletonApproximationsRadiiRoche1983}\footnote{Note that during the calculation of this grid, RL sizes are computed exactly to construct the quantity $r_\Psi/r_{\rm RL}$. These values are then interpolated by MESA with fractional RL sizes computed with the Eggleton approximation of $r_{\rm RL}$.}.
We do not model mass nor energy transfer in these models, as this is not needed for detached or equal mass binaries.

\subsection{Main sequence evolution of a single rotating star}\label{ssec:singlemodel}
We compare the main sequence evolution of a $5M_\odot$ star rotating initially at around 70\% critical velocity at Solar metallicity with and without the modified atmospheric BCs developed in Sect. \ref{sec:bound}. 
In MESA, the default BCs are integrated from a standard spherically symmetric, plane parallel, grey Eddington atmosphere \citep{eddingtonInternalConstitutionStars1926}. 
Our new BCs however take into account the modified pressure and area of the outer surface of the star, as represented by the correction factors $f_P, f_T$ and the non-spherical area $S_\Psi$ in Eqs. \eqref{eq:pressbound}-\eqref{eq:tempbound}.
\par
\begin{figure}
	\centering
	\includegraphics{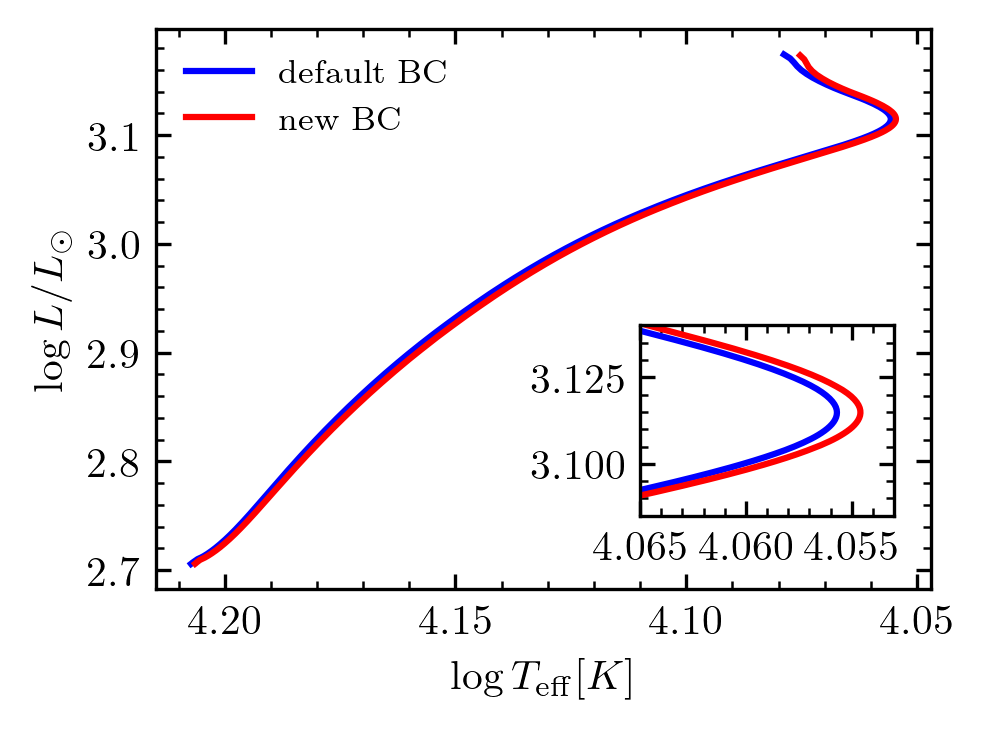}
	\caption{Hertzsprung-Russel diagram of the main sequence evolution of 5$M_\odot$ rotating stars in MESA using the default and new BCs. 
		The inset zooms in on the blue hook near the terminal age main sequence.}
	\label{fig:HRD_single}
\end{figure}
\begin{figure}
	\centering
	\includegraphics{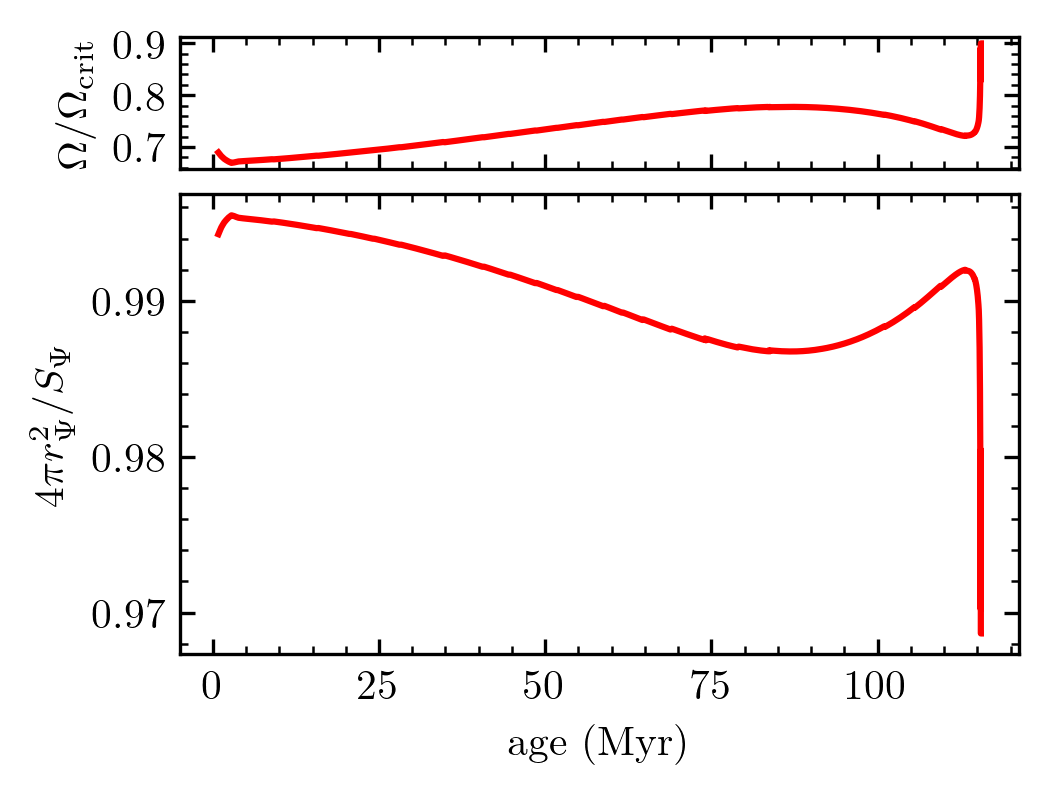}
	\caption{Fraction of equivalent spherical area to true area at the surface as function of age of the single rotating $5M_\odot$ model. 
		A lower fraction indicates a further departure from spherical symmetry due to rotation. 
		The top panel shows the fractional critical rotation rate for comparison.}
	\label{fig:areafrac_single}
\end{figure}

Figure \ref{fig:HRD_single} shows the evolutionary track of the single rotating stars in a Hertzsprung-Russel diagram.
We see that the change in BC has little impact on the overall evolution, with the largest difference of the surface properties occurring near hydrogen depletion ($X_c = 0.001$), where at constant $L$, the corresponding $T_{\rm eff}$ varies by about 0.7\%.
Since the spherical BC calculates $T_{\rm eff}$ from $L = 4\pi r^2T_{\rm eff}^4$, comparing with Eq. \eqref{eq:tempbound} at constant luminosity and radius results in the temperatures differing by a factor $\sqrt[4]{4\pi r_\Psi^2 / S_\Psi}$.
We plot the quantity $4\pi r_\Psi^2 / S_\Psi$ as function of stellar age in Fig. \ref{fig:areafrac_single}.
As the star loses thermodynamic equilibrium nearing hydrogen exhaustion, it contracts and tends toward critical rotation at the surface (recall we are enforcing solid body rotation).
At this limit, the area ratio reaches its minimum of about 0.945, which, per Eq. \eqref{eq:globalteff}, at constant luminosity, leads to an effective temperature ratio of 0.986, meaning a 2.4\% difference.
In the bulk of the main sequence however, we have an area ratio of around 0.99, which corresponds to just a 0.25\% difference in effective temperature.


\subsection{Evolution of a detached B-type star}\label{ssec:detachedmodel}
We modeled the evolution of a $5M_\odot$ solar metallicity B-type star in a binary with a point mass companion of $3M_\odot$ ($q=0.6$) with an initial period of 3 days until the star reaches its RL.
With this setup we mimic the situation of intermediate mass stars found in short-period eclipsing binaries.
We varied the rotation distortion model between (1) no rotation, (2) single star rotation where only centrifugal forces are taken into account and (3) tidal deformation that account for both centrifugal and tidal forces.
For the latter two, we also varied the atmospheric BCs between the new atmospheric BCs (Sect. \ref{sec:bound}) and the default spherical ones.
For the non rotating model, the new BCs reduce to the default ones, as there is no departure from spherical symmetry.
\par
\begin{figure}
	\centering
	\includegraphics{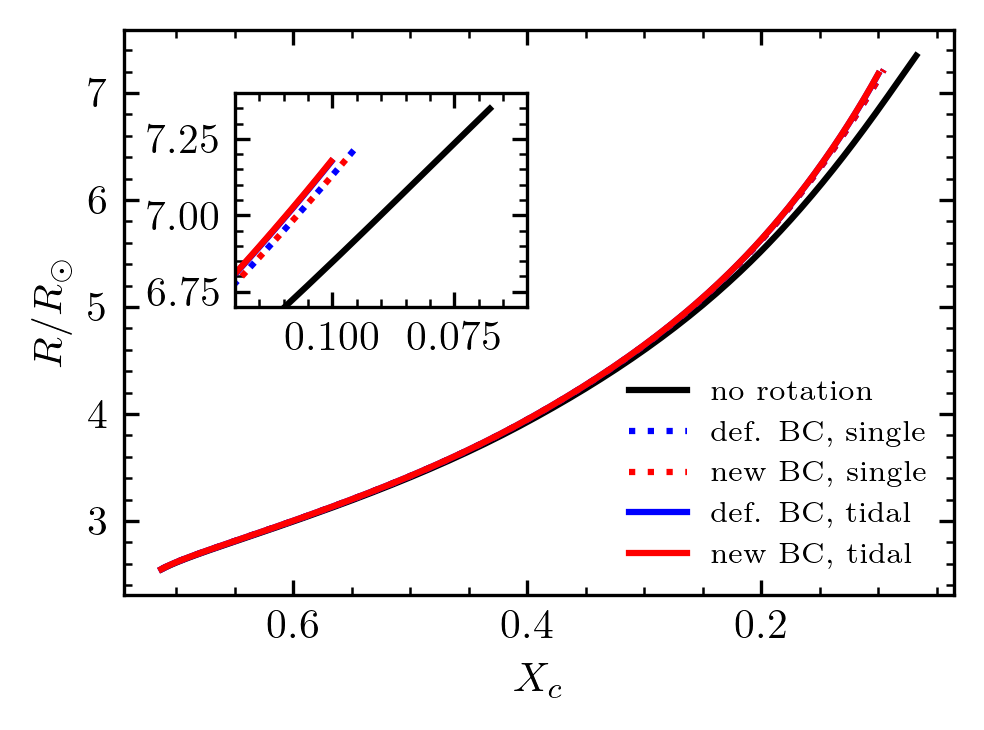}
	\caption{Radius evolution of a $5M_\odot$ star starting on a 3-day orbit around a $3M_\odot$ point mass, as function of central hydrogen fraction $X_c$.
		The difference between the rotation models is small while the one resulting from the different BCs is imperceivable.}
	\label{fig:r_xc_detached}
\end{figure}

Figure \ref{fig:r_xc_detached} shows the radius evolution of the stars as function of the central hydrogen fraction $X_c$.
At the onset of evolution, where the star is well within its RL, we see negligible difference between the various assumptions on BCs or rotation.
However, as the star approaches its RL size, at constant $X_c$ the family of rotating tracks differ from the non-rotating model by up to 5\%.
This is a significant difference, and is resolvable by very precise observations of eclipsing binaries, whose errors on measured radii can be better than 3\% \citep[][and references therein]{torresAccurateMassesRadii2010, graczykLatetypeEclipsingBinaries2018, pavlovskiPhysicalPropertiesCNO2018, tkachenkoMassDiscrepancyIntermediate2020}.
Therefore, modeling stars in eclipsing binaries as non-rotating in evolution codes could induce a systematic mismatch when fitting measured radii to inferred radii from such models.
Finally, we note that the difference in radii between the tidal and single rotating star models, which is at most 1\%, is only resolvable by the most accurate of observations, if at all.
Furthermore, the difference in radii between considered BC is on the order of 0.1\% and is not perceivable on the figure.

\subsection{Evolution of a twin overcontact binary}
We evolved a $32\,M_\odot$ star with a metallicity of $Z = Z_\odot / 2$ in a binary system with its twin, meaning all properties of the stars are identical.
In particular, this means that the mass ratio is unity, and we can ignore binary interaction effects like mass or energy transfer.
We start the evolution at the ZAMS with a 1.5 day orbit and stop the simulation once the stars reach the outer Lagrangian point.
This setup is the closest we can currently and consistently model massive overcontact binaries like VFTS 352 in the Large Magellanic Cloud \citep{almeidaDiscoveryMassiveOvercontact2015}.
\par
\begin{figure}
	\centering
	\includegraphics{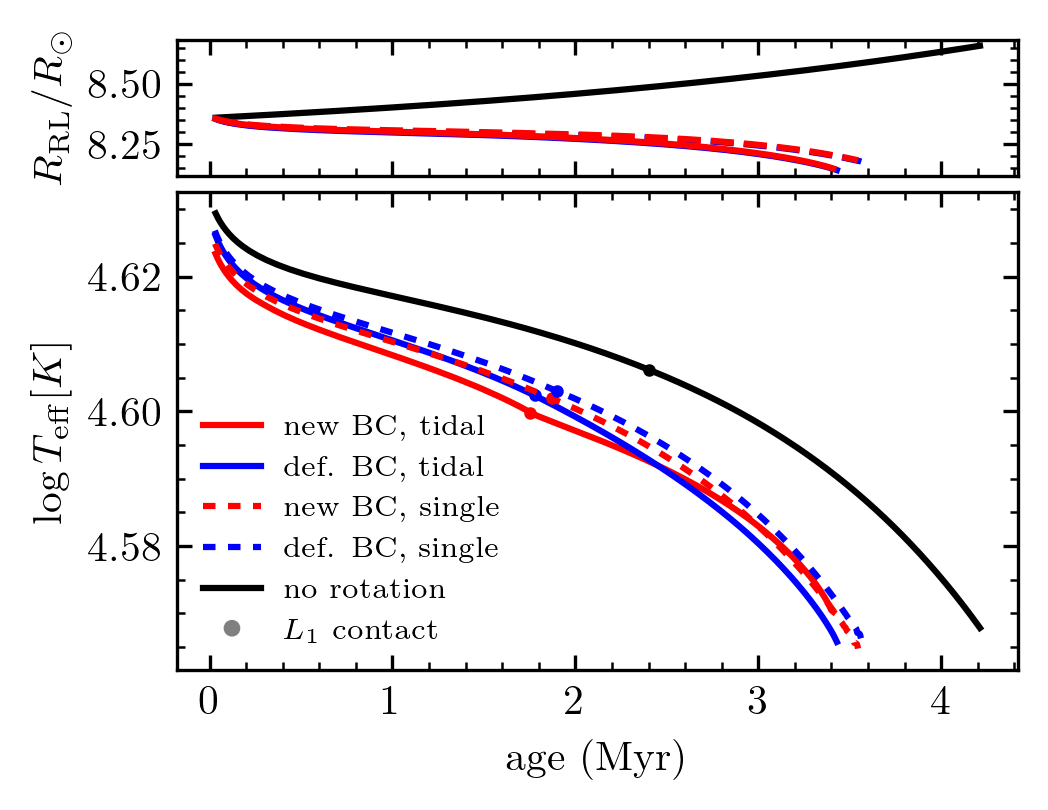}
	\caption{Effective temperature evolution of the $32\,M_\odot$ contact model with the different assumptions on BC and rotation model.
		We notice that the models including rotation are up to 5\% cooler during overcontact.
		Between the different rotation models and BCs, we find only minor differences.
		The top panel shows the RL evolution, which explains the difference in age at onset of overflow of the Lagrangian points through spin orbit coupling.
	}
	\label{fig:teff_age_contact}
\end{figure}

In Fig. \ref{fig:teff_age_contact} we show the effective temperature evolution of the $32\,M_\odot$ model under the various rotation model and BC assumptions.
We notice again, like in the detached case, that the observable stellar parameter (in this plot $T_{\rm eff}$) can vary up to 5\% in the late stages of the overcontact configuration.
The difference between the rotation models are minor, on the order of 1\%, and the tidally distorted model with the new BCs of Sect. \ref{sec:bound} shows a rather complex evolution.
It kinks at the RL filling point and subsequently crosses the other rotating tracks as the degree of overcontact increases.
This behavior can be related to the crossing of the structure correction factors $f_P$ and $f_T$ as $r_\Psi$ keeps rising, see Fig. \ref{fig:fp_r_super}.
Moreover, there is a $\unsim25\%$ age discrepancy between the rotating and non-rotating models.
This difference can be understood in the context of spin orbit coupling.
In the non-rotating model, the star can store no angular momentum, so that tidal synchronization has no effect (at least with the current implementation in MESA) and the RL size slowly increases due to wind mass loss (see top panel Fig. \ref{fig:teff_age_contact}).
However, when the star is modeled with angular momentum, tidal synchronization keeps the star rotating close to the orbital velocity at the cost of decreasing the orbital separation, thus reducing the RL size and accelerating the points of contact and $L_2$ overflow.
This effect is also visible in the detached binary of Sect. \ref{ssec:detachedmodel}.

\section{Discussion and conclusions}\label{sec:conc}
We developed the methodology needed to account for tidal deformation in binary stars in 1D stellar evolution codes.
They are represented in the structure correction factors $f_P$ and $f_T$ that multiply the equations of hydrostatic equilibrium and radiative energy transport, respectively.
Additionally, modified expressions of the Eddington limit and atmospheric boundary conditions as a result of the deformation have been calculated.
\par
We showed that the radii predicted by stellar evolution models can shift by $\unsim5\%$ if centrifugal deformation is included, while also including tidal deformation has a smaller $\unsim1\%$ effect.
This means comparing observed stellar parameters to non-rotating models could lead to a mismatch and has implications for high-precision astrophysics.
In a study where observed surface properties with uncertainties on the order of 5\% are used, centrifugal deformation has to be included in the stellar models.
If, in binaries, the precision of measurement is better than 1\%, then the tidal deformation has to be taken into account.
Note that the effect observed here is only due to geometrical considerations, that is, the equator bulges out due to rapid rotation or the presence of a companion.
Mixing as a result of rapid rotation was not included and will further affect the radius evolution.
Generally, the effect of rotational mixing is to keep the stars compact as fresh fuel is introduced into the core, so this may somewhat cancel the geometrical effect.
However, the exact interplay and outcome of these competing effects is far from established.
\par
The methods developed here have a wide range of applicability as they can be used for any (semi-)detached binary of arbitrary mass ratio.
This can be used, for instance, to study the effect of tidal forces in mass transfer stability, which can have a potential impact on the formation of gravitational wave sources through stable mass transfer \citep{vandenheuvelFormingShortperiodWolfRayet2017, marchantRoleMassTransfer2021}.
There are however limitations to our model, most notably that the structure correction factors computed from the binary Roche potential only apply to fully tidally locked systems.
The general case, where rotation of the components is free, cannot be described with a conservative potential, and is thus not suitable to the treatment of \citet{kippenhahnSimpleMethodSolution1970} and \citet{endalEvolutionRotatingStars1976} used here.
Developing a global model of free rotation including tides is a complex problem, but a potential workaround is to consider free rotation for stars well within their RLs and use the single star, centrifugal deformation corrections, and then implement a switch to the synchronized binary corrections once a star fills an appreciable amount of its RL.
\par
For contact configurations, currently only the mass ratio of unity can be considered consistently, as energy transfer in the overcontact layers is zero.
When we include the processes of mass and energy transfer in overcontact layers, which will be the subject of the following paper in this series, the methods we developed in this work enable the consistent modeling of all overcontact binary systems.

\begin{acknowledgements}
M.F. thanks the Flemish research foundation (FWO, Fonds voor Wetenshappelijk Onderzoek) PhD fellowship No. 11H2421N for its support.
P.M. acknowledges support from the FWO junior postdoctoral fellowship No. 12ZY520N.
The research leading to these results has received funding from the European Research Council (ERC) under the European Union's Horizon 2020 research and innovation programme (grant agreement numbers 772225: MULTIPLES).

\end{acknowledgements}
\begin{appendix}
\section{Polynomial fits in the rotating shell potential}\label{app:rotatingfits}
In the case of the rotating star potential \eqref{eq:potrotstar}, integral expressions can be written down to compute all relevant quantities required to use the modified stellar structure equations of Sect. \ref{ssec:equations}.
\citet{paxtonModulesExperimentsStellar2019} numerically integrated these and provided polynomial fits in the variable $\omega = \Omega/\Omega_{\rm crit}$, where $\Omega_{\rm crit} = \sqrt{GM/r_{\rm e}^3}$ is the classical critical velocity of equatorial radius $r_{\rm e}$.\par
To improve on the errors introduced by such fits, we recalculated the integrals and used the Levenberg-Marquardt non-linear least squares algorithm from \texttt{scipy} \citep{virtanenSciPyFundamentalAlgorithms2020} to find more accurate polynomial fits.
\par
First, for the equipotential volumes $V_\Psi$, we have the fit:
\begin{equation}
	V_\Psi(\omega) \simeq \frac{4\pi}{3}r_{\rm e}^3\left(1-\frac{\omega^2}{2}+0.1180\omega^4-0.07688\omega^6\right),
\end{equation}
with a maximal error of 0.20\% in $\omega \in [0, 1]$.
Note that since $V_\Psi(1)$ is an analytical result \citep{kopalCloseBinarySystems1959}, and the leading order term in the expansion is set by requiring that a slow rotating shell is ellipsoidal, this fit has only one free parameter.
For the volume equivalent radius $r_\Psi$ and equatorial radius $r_{\rm e}$ have the fits:
\begin{equation}
	r_\Psi(\omega) \simeq r_{\rm e}\left(1-\frac{\omega^2}{6}+0.02025\omega^4-0.03870\omega^6\right),
\end{equation}
with a maximal error of 0.11\% in $\omega \in [0, 1]$, and,
\begin{equation}\label{eq:polyequarad}
	r_{\rm e} \simeq r_\Psi(\omega) \left(1+\frac{\omega^2}{6}-0.005124\omega^4+0.06562\omega^6\right),
\end{equation}
with a maximal error of 0.15\% in $\omega \in [0, 1]$.
Next, the surface area $S_\Psi$ and average effective gravity $\avg{g}$ are approximated by:
\begin{align}
	S_\Psi \simeq {}&4\pi r_{\rm e}^2 \left(1-\frac{\omega^2}{3}+0.08696\omega^4-0.05079\omega^6\right),\\
	S_\Psi\avg{g} \simeq {}&
	4\pi GM_\Psi\times\notag\\
	&\left(1-\frac{2}{3}\omega^2+0.2989\omega^4+0.006978\omega^6\right),
\end{align}
with a maximal error of 0.07\% and 0.25\% in $\omega \in [0, 1]$, respectively.
Since the inverse gravity diverges when $\omega \rightarrow 1$, we require a non-polynomial term to represent it appropriately in that limit.
Therefore, constructing the auxiliary functions:
\begin{align}
	A(\omega) = 1 &+ 0.3293\omega^4-0.4926\omega^6-0.5560\ln(1-\omega^{5.626}),\\
	B(\omega) = 1 &+ \frac{\omega^2}{5} + 0.4140\omega^4 - 0.8650\omega^6 \notag\\
	&- \frac{3}{2}\times 0.5580\ln(1 - \omega^{5.626}),
\end{align}
we have then for the average inverse effective gravity $\avg{g^{-1}}$ and the specific moment of inertia $i_{\rm rot}$ the fits
\begin{align}
	S_\Psi \avg{g^{-1}} &\simeq \frac{4\pi r_{\rm e}^4}{GM_\Psi}A(\omega)\\
	i_{\rm rot}(\omega) &\simeq \frac{2}{3}r_{\rm e}^2 \frac{B(\omega)}{A(\omega)},
\end{align}
with a maximal error of 0.85\% and 0.68\% in $\omega \in [0, 0.9999]$, respectively.
\par
For the correction factors $f_P$ en $f_T$ we construct the fits:
\begin{align}
	f_P(\omega) &\simeq \frac{1-\frac{2}{3}\omega^2+0.2133\omega^4-0.1068\omega^6}{A(\omega)},\\
	f_T(\omega) &\simeq \frac{1-0.07955\omega^4-0.2322\omega^6}{A(\omega)},
\end{align}
with a maximal error of 0.56\% and 0.58\% in $\omega \in [0, 0.9999]$, respectively.
\par
Finally, with the auxiliary function:
\begin{align}
	C(\omega) = 1 &+ \frac{17}{60}\omega^2 + 0.4010\omega^4 - 0.8606 \omega^6 \notag\\
	&- 0.9305\ln(1-\omega^{5.626}),
\end{align}
the implicit equation from which $\omega$ can be obtained from $j_{\rm rot}$ is:
\begin{equation}
	\frac{j_{\rm rot}}{\sqrt{GM_\Psi r_\Psi}} \simeq \frac{2}{3}\frac{\omega C(\omega)}{A(\omega)},
\end{equation}
with a maximal error of 0.54\% in $\omega \in [0, 0.9999]$.
\par
With these fits, we improve all the fits of \citet{paxtonModulesExperimentsStellar2019}, and improve especially the maximal error of $f_T$, namely from 1.6\% to 0.58\%.

\section{Matching the tidal and single rotating star distortion models}\label{app:rotating-roche}
In Fig. \ref{fig:fp_r_super}, we make the comparison of the structure correction factor $f_P$ of the tidal model as well as the single rotating star model in one plot.
This is meant to illustrate the corrections that would be used in a simulation with the MESA code if rotation is included in a model for a binary system without including tidal deformation.
In practice, a tidally distorted star does not have a well defined equatorial radius with which to define $\omega$, but nevertheless the current implementation of MESA makes use of the equipotential radius $r_\Psi$, the mass contained within the shell $M_\Psi$ and the angular frequency of the shell $\Omega_\Psi$ to derive the corrections as if these properties corresponded to a single rotating star with cylindrical symmetry.
\par

For a tidally synchronized binary, the angular velocity is derived from Kepler's third law:
\begin{equation}
	\Omega = \sqrt{\frac{GM_1(1+q)}{a^3}},
\end{equation}
with $q=M_2/M_1$ the mass ratio and $a$ the separation of the binary.
Next we consider the equivalent single rotating star.
Given the mass $M_1$ and equatorial radius $r_{\rm e}$, it has a critical rotation rate:
\begin{equation}
	\Omega_{\rm crit} = \sqrt{\frac{GM_1}{r_{\rm e}^3}}.
\end{equation}
Dividing these equations, and setting $\omega = \Omega/\Omega_{\rm crit}$ as the dimensionless rotation rate, we arrive at an implicit equation for $\omega$:
\begin{equation}\label{eq:implicit}
	\omega^2 = \left(\frac{r_e(\omega, r_\Psi)}{a}\right)^3(1+q) = \left(\frac{r_\Psi}{a}f_{r_{\rm e}}(\omega)\right)^3(1+q).
\end{equation}
where $f_{r_{\rm e}}$ relates the volume equivalent radius to the equatorial radius as function of the dimensionless rotation rate in a single rotating star.
For a given value of $r_\Psi/a$ and $q$, this implicit equation gives the value of $\omega$ that a MESA simulation would determine to compute the corrections to the stellar structure equations, despite there not being a well defined equatorial radius.
The values shown in Fig. \ref{fig:fp_r_super} for the single star equivalent correspond to evaluating the results of Appendix \ref{app:rotatingfits} for $f_P(\omega)$, with $\omega$ derived from $r_\Psi/a$ and $q$ with the implicit Eq. \eqref{eq:implicit}.
\end{appendix}
\bibliographystyle{bibtex/aa}
\bibliography{bibtex/Massive_Stars.bib}

\end{document}